\documentclass[12pt]{article}
\usepackage{graphicx}

\usepackage{fancyhdr}
\usepackage{subfig}%
\usepackage{xspace}%
\usepackage{multirow}%
\usepackage{dcolumn} 

\usepackage[table,usenames,dvipsnames]{xcolor}

\usepackage{amsmath,amssymb}

\usepackage{topcapt}

\newcommand\pubdate{\today}

\def\napoli{Division of Physics, Mathematics and Astronomy\\
California Institute of Technology, 91125 Pasadena, California, USA}
\def\support{\footnote{Work supported by funding from California Institute of Technology High En- ergy Physics under Contract DE-SC0011925 with the United States Department of Energy.}}

\def\Title#1{\begin{center} {\Large #1 } \end{center}}
\def\Author#1{\begin{center}{ \sc #1} \end{center}}
\def\Address#1{\begin{center}{ \it #1} \end{center}}

\newcommand\pubblock{\rightline{\begin{tabular}{l} \pubdate  \end{tabular}}}
\newenvironment{Abstract}{\begin{quotation}  }{\end{quotation}}
\newenvironment{Presented}{\begin{quotation} \begin{center} 
             PRESENTED AT\end{center}\bigskip 
      \begin{center}\begin{large}}{\end{large}\end{center} \end{quotation}}

\def\beq{\begin{equation}}
\def\eeq#1{\label{#1}\end{equation}}
\def\eeqn{\end{equation}}

\def\beqa{\begin{eqnarray}}
\def\eeqa#1{\label{#1}\end{eqnarray}}
\def\eeqan{\end{eqnarray}}

\let\bar=\overbar

\def\Dslash{\not{\hbox{\kern-4pt $D$}}}
\def\dslash{\not{\hbox{\kern-2pt $\del$}}}

\def\msb{{\bar{\ssstyle M \kern -1pt S}}}

\newcommand{\GEANT} {{\textsc{geant}}\xspace}

\newcommand{\GeV}{\ensuremath{\,\text{Ge\hspace{-.08em}V}}\xspace}

\newcommand{\PTm}{\ensuremath{{p}_\mathrm{T}\hspace{-1.02em}/\kern 0.5em}\xspace}

\newcommand{\ETm}{\ensuremath{E_{\mathrm{T}}^{\text{miss}}}\xspace}
\newcommand{\MET}{\ETm}

\newcommand{\ETslash}{\ensuremath{E_{\mathrm{T}}\hspace{-1.1em}/\kern0.45em}\xspace}

\newcommand{\ptvecmiss}{\ensuremath{{\vec p}_{\mathrm{T}}^{\kern1pt\text{miss}}}\xspace}

\newcommand{\cPgg}{\ensuremath{\gamma}} 
\newcommand{\cPqb}{\ensuremath{\mathrm{\textbf{b}}}} 
\newcommand{\cPaqb}{\ensuremath{\overline{\mathrm{\textbf{b}}}}} 
\providecommand{\PH}{\ensuremath{\mathrm{\textbf{H}}}\xspace} 

\providecommand{\Pgt}{\ensuremath{\tau}\xspace} 

\newcommand{\HGG}{\ensuremath{\mathrm{H}\to\gamma\gamma}}

\newcommand{\rpv}{\ensuremath{\rlap{\kern.2em/}R}\xspace}

\begin{document}
\begin{titlepage}
\pubblock

\vfill
\Title{CMS precision timing physics impact for the HL-LHC upgrade}
\vfill
\Author{Olmo Cerri\support, on the behalf of the CMS collaboration.}
\Address{\napoli}
\vfill
\begin{Abstract}
As part of the Phase II upgrade program, the Compact Muon Solenoid (CMS) detector will incorporate a new timing layer designed to measure minimum ionizing particles (MIPs) with a time resolution of $\sim$30 ps. Precision timing will mitigate the impact of the challenging levels of pileup expected at the High Luminosity LHC. The time information assigned to each track will enable the use of 4D-vertexing which will render a 5-fold pile-up reduction, thus recovering the current conditions. Precision timing will also enable new time-based isolations and improved $b$-tagging algorithms. All of this translates into a $\sim20\%$ gain in effective luminosity when looking at di-Higgs boson events decaying to a pair of $b$-quarks and two photons. We present the expected improvements in physics performance with precision timing with the upgraded CMS detector.
\end{Abstract}
\vfill
\begin{Presented}
Thirteenth Conference on the Intersections of Particle and Nuclear Physics\\
Palm Springs, CA, USA,  May 29 - June 3, 2018
\end{Presented}
\vfill
\end{titlepage}
\def\thefootnote{\fnsymbol{footnote}}
\setcounter{footnote}{0}

\section{Introduction}

The primary goal of the Phase-2 upgrade for the High-Luminosity LHC (HL-LHC) is to maintain the excellent performance of the CMS detector in efficiency, resolution, and background rejection for all final state particles and physical quantities used in data analyses.
The CMS Upgrade Technical Proposal~\cite{Butler:2020886} presents, and the Scope Document~\cite{Butler:2055167} further specifies, a detailed upgrade plan to deploy an improved CMS detector by 2026, at the start of the HL-LHC operation~\cite{Apollinari:2017cqg}.
It identifies changes necessary to withstand radiation damage effects and describes upgrades of the CMS components needed to overcome the challenge posed by the high rate of concurrent collisions per beam crossing (pileup) at the HL-LHC.

The LHC will operate at a stable luminosity of 5.0$\times 10^{34}$~cm$^{-2}$s$^{-1}$, yielding 140 pileup collisions by continuously tuning the beam focus and the crossing profile during a fill.
An ultimate scenario, with 7.5$\times 10^{34}$~cm$^{-2}$s$^{-1}$ luminosity and 200 pileup collisions per beam crossing, would provide 40\% more accumulated data.
At 140 or 200 pileup collisions, a hard interaction, one that probes energy scales of order 0.1--1~TeV, occurs in less than 1\%  of the total number of interactions simultaneously recorded by the detector.
The spatial overlap of tracks and energy deposits from these collisions can degrade the identification and the reconstruction of the hard interaction, and can increase the rate of false triggers.

Pileup mitigation in CMS relies upon particle-flow event reconstruction~\cite{CMS:2009nxa}, which removes from relevant quantities charged tracks inconsistent with the vertex of interest, and neutral deposits in the calorimeters with ansatz-based statistical inference techniques like PUPPI~\cite{Bertolini:2014bba}.
The high spatial granularity of the subdetectors will enable the upgraded CMS detector to separate vertices, to identify the hard collision, and to measure signal particles with good efficiency in the offline analyses~\cite{Butler:2020886, Butler:2055167}. However, in the transition from 140 to 200 pileup events,  the probabilities of spatial overlap grow in all subdetectors, and subsequently particle-flow algorithms begin to fail at a substantial rate.
The resulting degradation in resolutions, efficiencies, and misidentification rates at 200 pileup events impacts several physics measurements~\cite{Butler:2055167, DP2016_065}.

The timing upgrade of the CMS detector will improve the particle-flow performance at high pileup to a level comparable to the Phase-1 CMS detector, exploiting the additional information provided by the precision timing of both tracks and energy deposits in the calorimeters.

\section{MIP Timing Detector}
\label{sec:MTDoverview}
\begin{figure*}[htb!]
\centering
\includegraphics[width=0.68\textwidth]{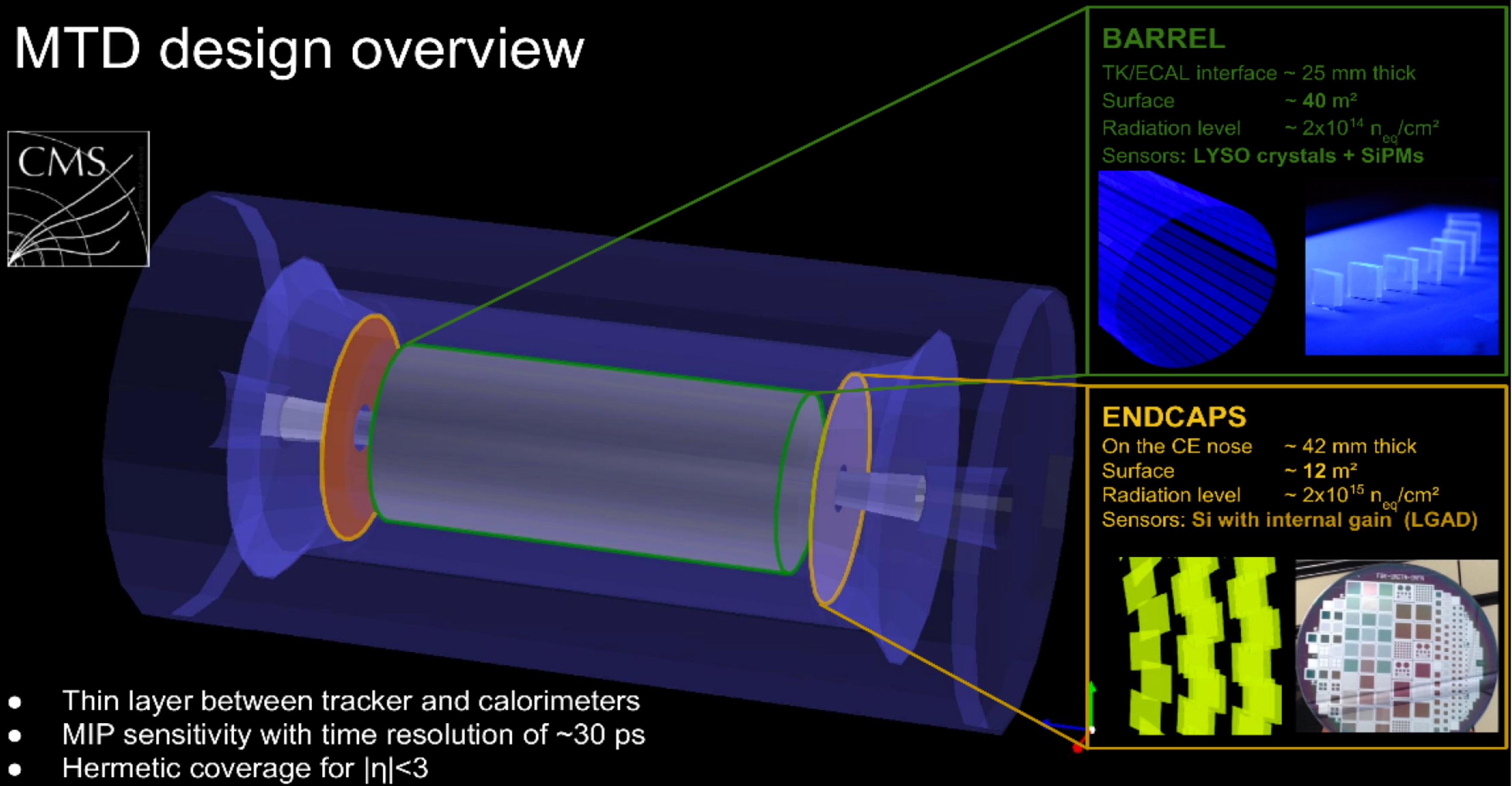}
\caption{
A simplified \GEANT geometry of the timing layer implemented in CMSSW for simulation studies comprises a LYSO barrel (grey cylinder), at the interface between the Tracker and the ECAL, and two silicon endcap (orange discs) timing layers in front of the CE calorimeter.
}
\label{fig:EvtDisplay}
\end{figure*}

The MTD will consist of a single layer device between the Tracker and calorimeters, covering up to $|\eta|\sim3$, with time resolution of order 30--40~ps for charged tracks throughout the detector lifetime.
Figure~\ref{fig:EvtDisplay} shows a simplified implementation in \GEANT of the proposed layout integrated in the CMS detector.
The MTD will comprise a barrel and an endcap region, with different technologies based on different performance, radiation, mechanics and schedule requirements and constraints.\\
The barrel timing layer will cover the pseudorapidity region up to $|\eta| = 1.48$.
After several dedicated beam tests, LYSO crystal scintillators read out with silicon photomultipliers (SiPMs)~\cite{gundacker2013time, LYSONIM, Anderson:2015tia} and silicon sensors with internal gain~\cite{White:2014oga,Pellegrini201412, Cartiglia:2015iua} emerged as a mature technology for the barrel timing layer.
Both LYSO based scintillators and SiPMs devices are mature technologies, with production and assembly procedures well established and standardized in industry.
The R\&D for a precision timing application is well advanced, and small prototypes consisting of LYSO:Ce crystals read out with SiPMs have been proven capable of achieving time resolution below 30~ps.
Both the crystals and the SiPM are proven to be radiation tolerant up to a neutron equivalent fluence of $2\times10^{14}$~cm$^{-2}$, when cooled to below $240$~$^{\circ}$K, equivalent to the integrated radiation at the end of life in the HL-LHC environment.\\
The endcap region can be instrumented with a hermetic, single layer of MIP-sensitive silicon devices with high time resolution, with a pseudorapidity acceptance from about $|\eta|=1.6$ to $|\eta|=2.9$.
Beam tests shown that silicon sensors with internal gain~\cite{White:2014oga,Pellegrini201412, Cartiglia:2015iua} are viable technology for the endcap timing layer, since the technology selected for the barrel cannot be extended to the endcap, due to radiation tolerance limitations.

\section{Impact of precision timing on the HL-LHC physics program}
In the time domain, pileup collisions at the HL-LHC will occur with an RMS spread of approximately 180--200~ps, constant during the fill and uncorrelated with the line spread along the beam line.
Slicing the beam spot in consecutive 30~ps exposures effectively reduces the number of vertices down to current LHC conditions, thereby recovering the Phase-1 quality of event reconstruction.
The essential basis for the proposed concept is that the time information from charged tracks is exploited in a space-time reconstruction of tracks and vertices.
In addition to the enhanced timing capabilities of the  calorimeters~\cite{Butler:2020886}, this approach requires a dedicated detector for precision timing of minimum ionizing particles (MIPs): the MIP timing detector (MTD).
\begin{figure}[t!]
\centering
\raisebox{-0.45\height}{\includegraphics[width=0.6\textwidth]{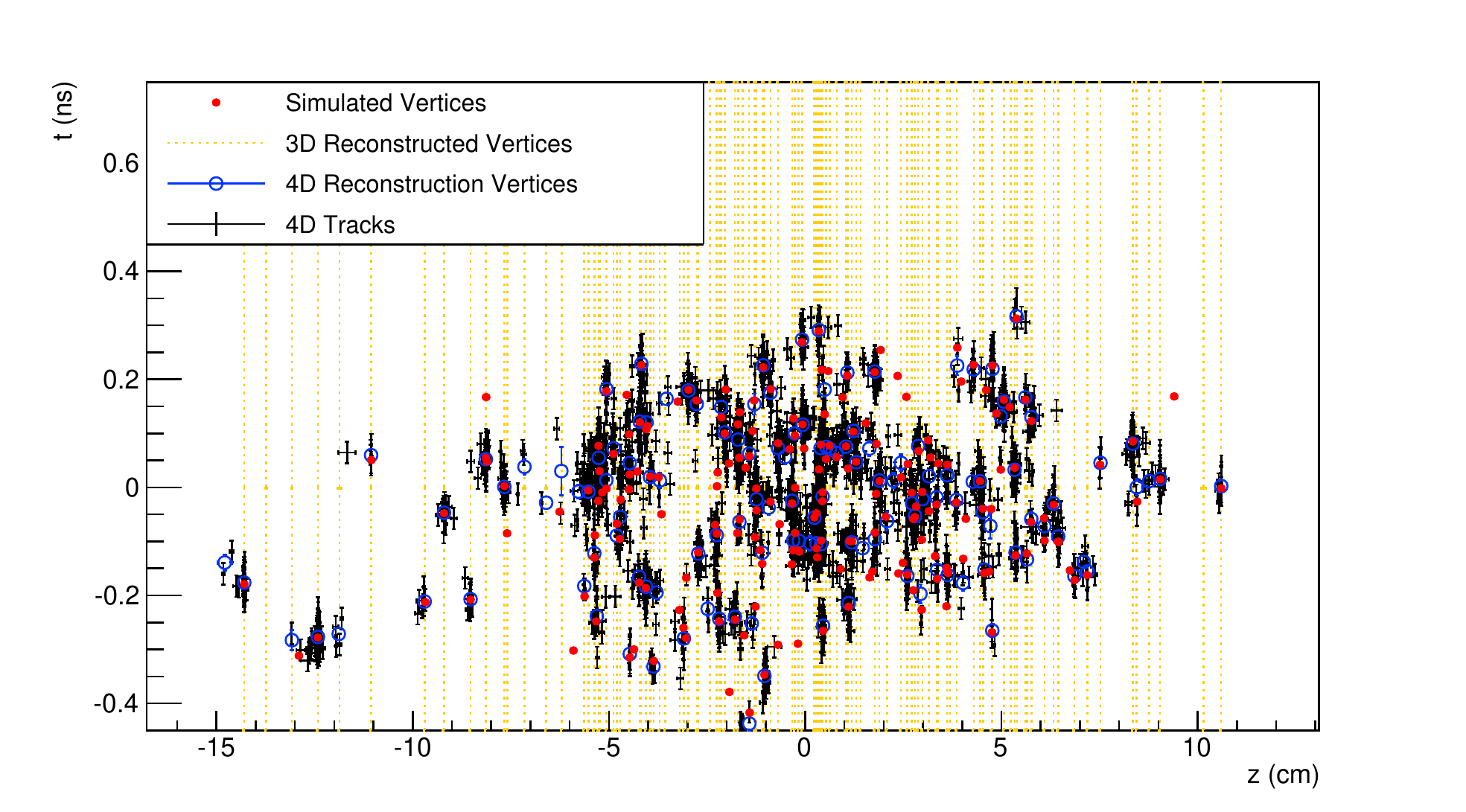}}
\raisebox{-0.5\height}{\includegraphics[width=0.35\textwidth]{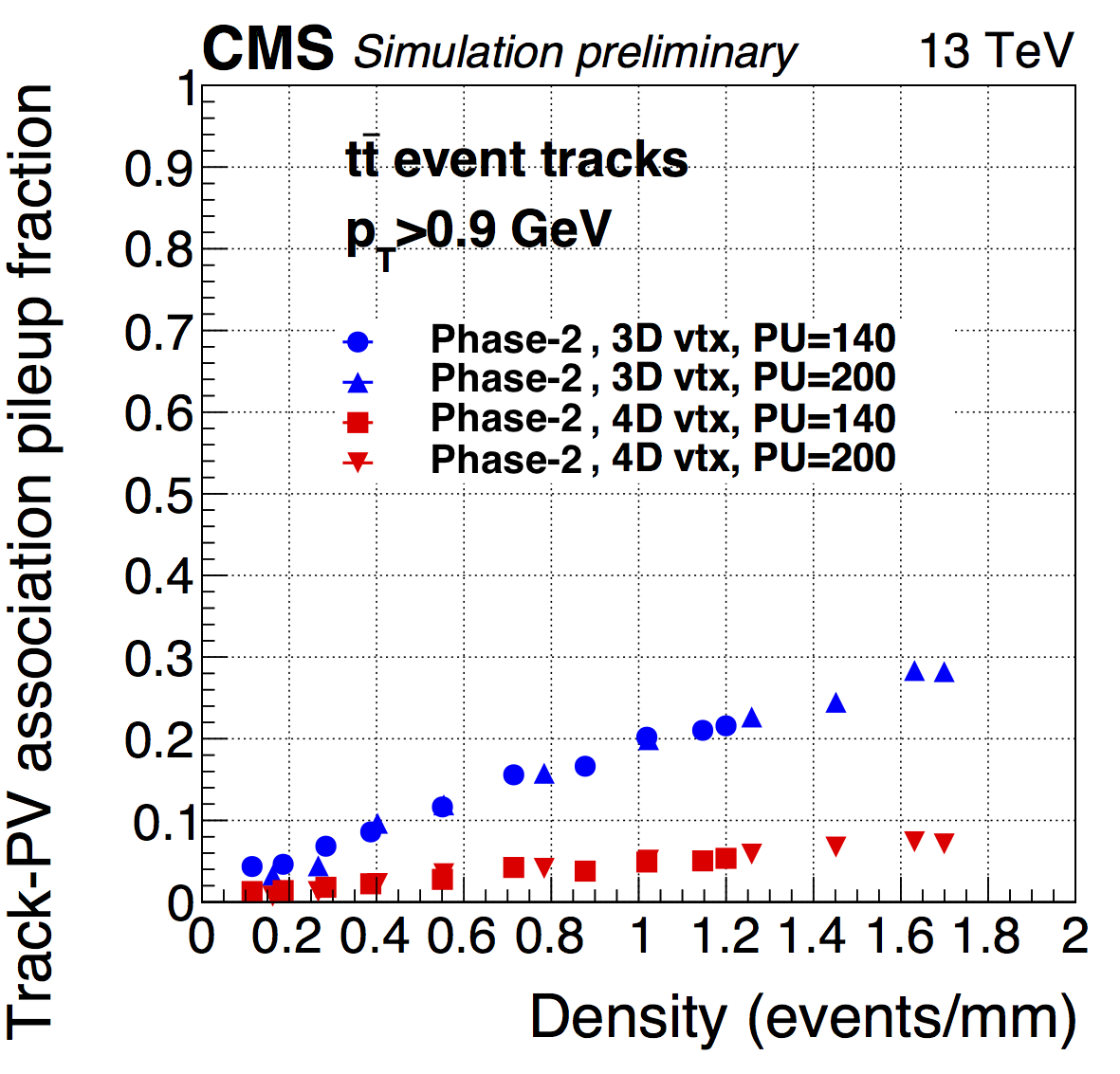}}
\caption{Left: Simulated and reconstructed vertices in a 200 pileup event assuming a MIP timing detector covering the barrel and endcaps. The vertical lines indicate 3D-reconstructed vertices, with instances of vertex merging visible throughout the event display. Right: Rate of tracks from pileup vertices incorrectly associated with the primary vertex of the hard interaction normalized to the total number of tracks in the vertex.
}
\label{fig:4Dvertex_200}
\end{figure}
The event display in Fig.~\ref{fig:4Dvertex_200} (left panel) visually demonstrates the power of space-time reconstruction in 200 pileup collisions, using a time-aware extension (4D) of the deterministic annealing technique adopted in vertex reconstruction by the CMS experiment~\cite{Chatrchyan:2014fea}.
According to simulation, instances of vertex merging are reduced from 15\% in space to 1\% in space-time.
Another quantitative measure of the performance improvement is shown in the right panel of Fig.~\ref{fig:4Dvertex_200}, showing the rate of tracks from pileup vertices incorrectly associated with the hard interaction vertex as a function of the line density of vertices.
The addition of track-time information with 30~ps precision reduces the wrong associations to a level comparable to that of the current LHC (vertex density of to about 0.3~mm$^{-1}$).
The performance of b-jet identification, which relies on vertex reconstruction, is enhanced.
The removal of pileup tracks from the isolation cones improves the identification efficiency for isolated leptons and photons, which are key signatures of many processes of interest for the HL-LHC program.
Similarly, the reconstruction of spatially extended objects and global event quantities that are vulnerable to pileup, such as jets and \MET{}, is also significantly improved.
At 200 pileup, the \MET{} resolution improves by about 10\% and the rate of reconstructed jets that are spuriously clustered particles from pileup interactions (``pileup jets") is reduced by up to 40\%, using track-time information in jet reconstruction.
All of this consistently motivates a precision timing detector in the barrel and in the endcaps, with about 30~ps resolution in order to enhance the HL-LHC physics reach.

\begin{figure}[hbtp]
\centering
\raisebox{-0.5\height}{\includegraphics[width=0.36\textwidth]{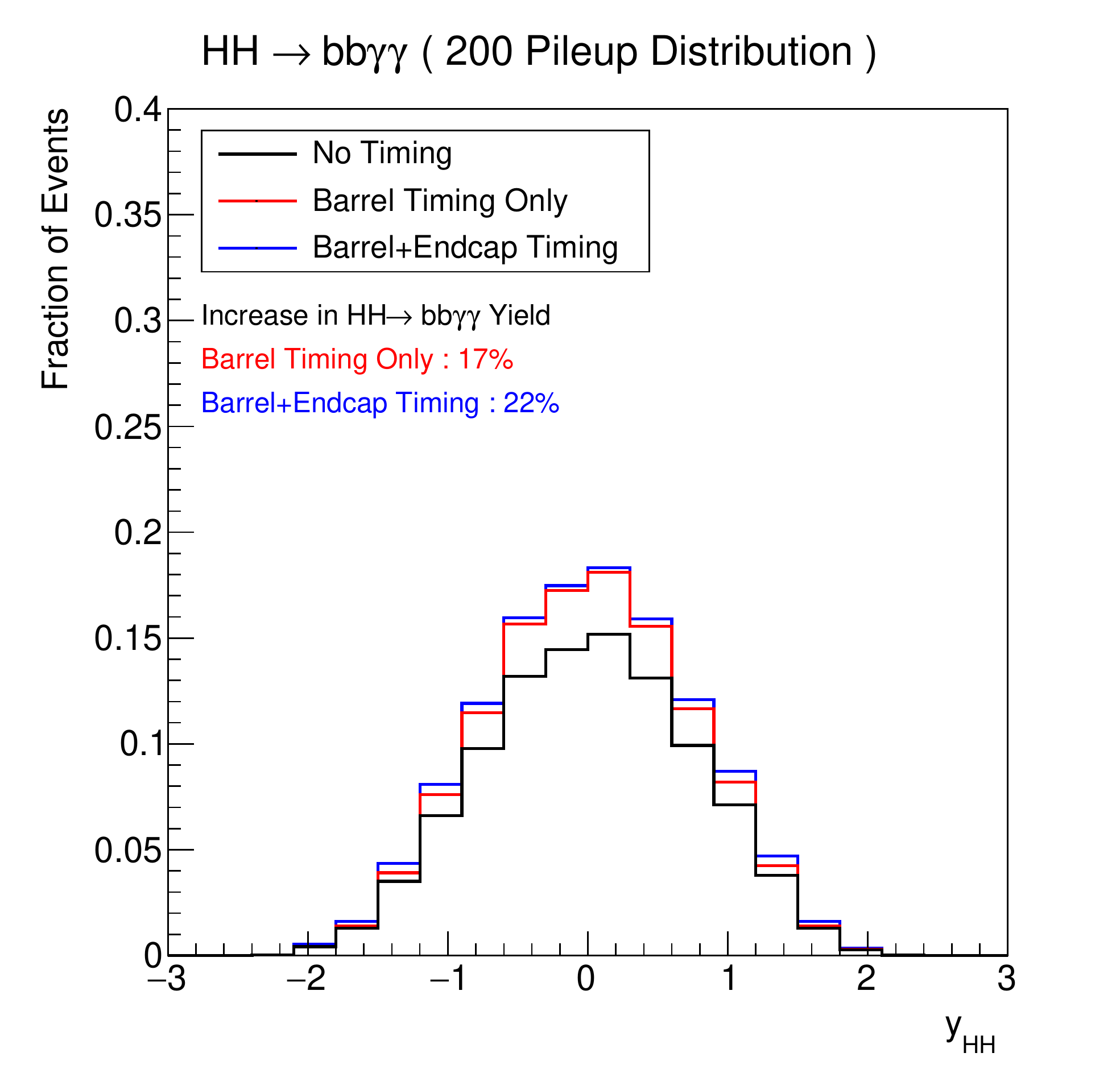}}
\raisebox{-0.5\height}{\includegraphics[width=0.36\textwidth]{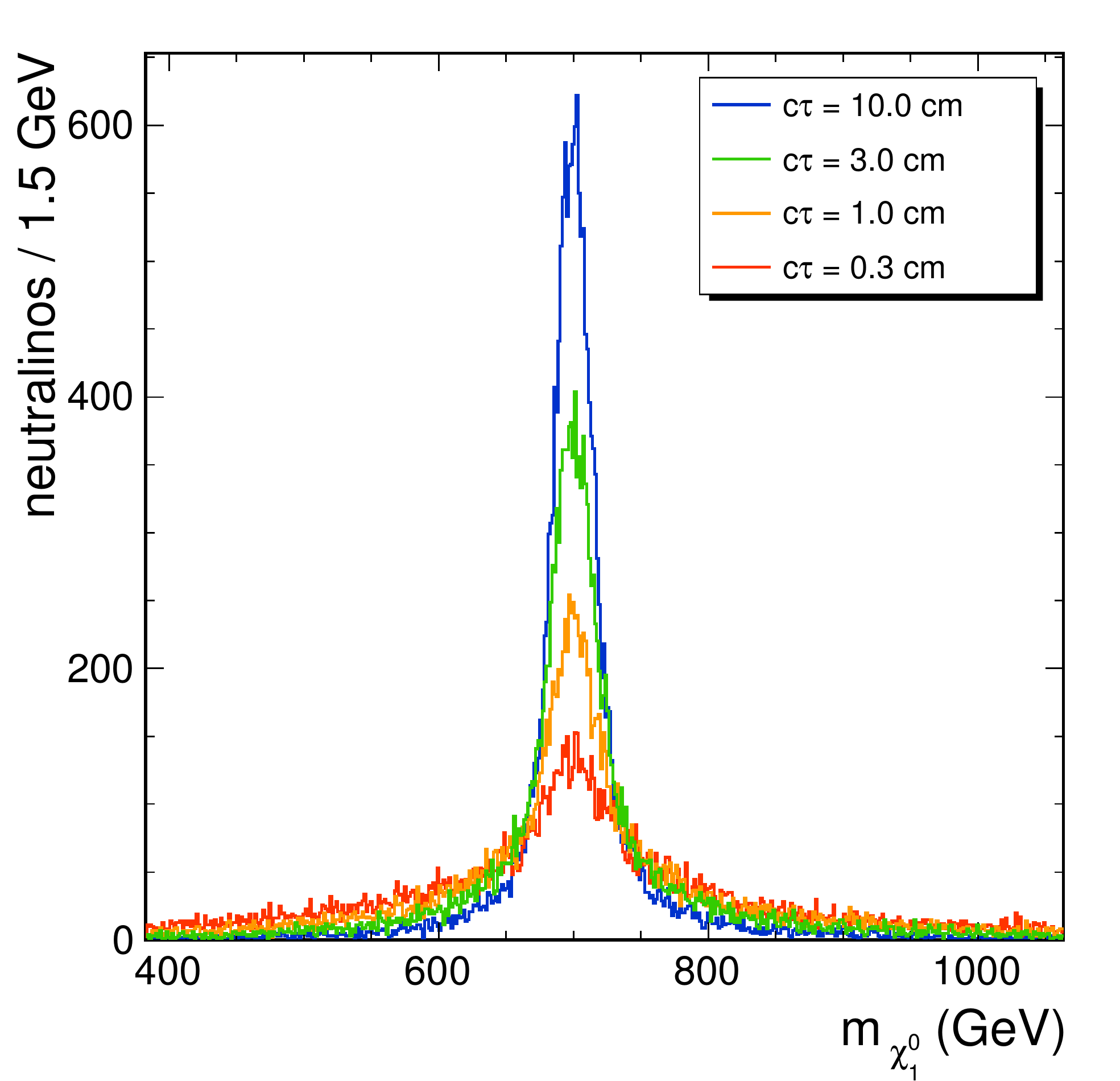}}
\caption{Impact on signal efficiency for $\PH\PH\to\cPqb\cPaqb\cPgg\cPgg$ for no-timing and two timing implementation
scenarios (left). Mass peak of a 700 \GeV\ neutralino reconstructed from the kinematic closure of the secondary vertex using time information with 30~ps resolution (right). }
\label{fig:HHAndllpSUSY}
\end{figure}

The CMS physics program at the HL-LHC will target a very wide range of measurements, including in-depth studies of the Higgs boson properties and direct searches for physics beyond the standard model (BSM).
The added value of a timing detector, quantified in terms of improved vertex identification, acceptance extension for isolated objects, improved \MET{} resolution, and pileup jet rate reduction, makes a significant impact on the CMS physics program across several channels.
A synopsis is presented in Table~\ref{table:impact}, where detector requirements are mapped into analysis and physics impacts.
\begin{table}
\centering
\caption{Representative signals for Higgs boson measurements and SUSY searches used to map each specific detector requirement into the relative performance gain at the analysis level (analysis impact) and in the measured physical quantity (physics impact).}
\label{table:impact}

\scriptsize
 \begin{tabular}{|l|l|l|l|} \hline
 Signal  & Detector requirement & Analysis impact &  Physics impact \\ \hline
\parbox[t]{1.8cm}{\HGG{}} &
 \parbox[t]{4.9cm}{30 ps photon and track timing  \\
                           $\bullet$ barrel: central signal \\
                           $\bullet$ endcap: improved \\ time-zero and acceptance} &
 \parbox[t]{4.8cm}{$S/\sqrt{B}$ : \\
 ~~$+20\%$ - isolation efficiency \\
 ~~$+30\%$ - diphoton vertex} &
 \parbox[t]{2.8cm}{$+25\%$ (statistical) precision on \\ cross section} \\ \hline
 \parbox[t]{1.8cm}{VBF+\\ $\PH$ $\to$ $\Pgt\Pgt$} &
 \parbox[t]{4.9cm}{30 ps track timing \\
                           $\bullet$ barrel: central signature \\
                           $\bullet$ endcap: forward jet tagging \\
                           $\bullet$ hermetic coverage: optimal \\ \MET{} reconstruction } &
 \parbox[t]{4.8cm}{$S/\sqrt{B}$ : \\
 ~~$+30\%$ - isolation efficiency \\
 ~~$+30\%$ - VBF tagging \\
 ~~$+10\%$ - mass (\MET{}) resolution} &
 \parbox[t]{2.8cm}{$+20\%$ (statistical) precision on \\ cross section \\ (upper limit \\ or significance)} \\ \hline
\parbox[t]{1.8cm}{$\PH\PH$} &
\parbox[t]{4.9cm}{30 ps track timing  \\
            $\bullet$ hermetic coverage} &
\parbox[t]{4.8cm}{signal acceptance : $+20\%$ \\
   ~~ b-jets and isolation efficiency} &
  \parbox[t]{2.8cm}{Consolidate \\ $\PH\PH$ searches} \\ \hline
 \parbox[t]{1.8cm}{$\chi^{\pm}\chi^0 \to$ \\ W$^\pm$H+\MET{}} &
 \parbox[t]{4.9cm}{30 ps track timing \\
                           $\bullet$ hermetic coverage: \MET{} } &
 \parbox[t]{4.8cm}{$S/\sqrt{B}$ : \\
 ~~$+40\%$ - reduction of \MET{} tails} &
 \parbox[t]{2.8cm}{$+150$~\GeV \\ mass reach} \\ \hline
 \parbox[t]{1.8cm}{Long-lived \\ particles} &
 \parbox[t]{4.9cm}{30 ps track timing \\
                           $\bullet$ barrel: central signature} &
 \parbox[t]{4.8cm}{mass reconstruction \\ of the decay particle} &
 \parbox[t]{2.8cm}{unique sensitivity \\ to split-SUSY and \\ SUSY with compressed spectra} \\ \hline
 \end{tabular}
 \end{table}
The {characterization of the Higgs boson properties}, with precision measurements of the Higgs boson couplings to standard model (SM) particles, and the search for rare SM and BSM decays, will benefit from the improved acceptance for isolated objects, and in the case of \HGG{} decays from improved vertex identification.
The quality of the isolation discriminant relies on the removal of pileup contributions close in angle to the candidate signal particle.
Therefore the efficiency gain is maximal with hermetic coverage, while barrel coverage is especially relevant for processes with central signatures.
Another crucial measurement at the HL-LHC is di-Higgs production, and consequently the direct measurement of the Higgs self-coupling. In this case, precision timing increases the signal yields for constant background in $\PH\PH\to\cPqb\cPaqb\cPgg\cPgg$ by 17\% from the barrel alone, and 22\% with hermetic coverage (Fig.~\ref{fig:HHAndllpSUSY}).
Similar enhancements are predicted for other important Higgs boson signatures, ranging from 15--20\% for HH$\to 4$b to 20--26\% for H$\to4\mu$, for constant background.
These acceptance extensions will provide improved precision in the measurement of rare decay modes and of statistically limited differential distributions, with sensitivity to Higgs boson pseudo-observables~\cite{deFlorian:2016spz}.
In the case of H$\to\tau\tau$ in the vector boson fusion (VBF) production mode, additional substantial gains arise from the improved quality of the \MET{} reconstruction and of the VBF jet identification.

The sensitivity to several {searches for new phenomena} is largely driven by the \MET{} resolution, which determines the background level for several BSM signatures.
The gain in the \MET{} resolution with track timing leads to a reduction of $\sim 40$\% in the tail of the \MET{} distribution above 130~\GeV, which approximately offsets the performance degradation for SUSY searches in the transition from 140 to 200 pileup~\cite{Butler:2055167}.
Additional benefits of the precision timing are anticipated in  searches with multi-lepton final states due to the increased efficiency of the lepton isolation selection, and in signatures where a direct measurement of the time of flight (TOF) of heavy particles is exploited.
For example, a TOF measurement in a detector in front of the calorimeters will reduce the model dependence in searches for Heavy Charged Stable Particles, now limited to particles that have little interactions with the calorimeters~\cite{CMS-PAS-EXO-14-007, Cerri:2018rkm}.
Moreover, the track-time reconstruction opens a new avenue in searches for neutral long lived particles (LLPs), postulated in many extensions of the standard model like Split-SUSY, GMSB, RPV SUSY, Stealth SUSY, SUSY models with compressed mass spectra and many others discussed in~\cite{PhysRevD.94.113003} and references therein.
The space-time information associated to the displaced decay vertex, will enable the kinematic reconstruction: for example, the direct measurement of the LLP mass (Fig.~\ref{fig:HHAndllpSUSY}), thus boosting the sensitivity of such searches and providing a novel method to characterize any future discovery.

\section{Summary and Future Considerations}
The addition of a precise timing detector, able to efficiently
tag charged particles, will significantly suppresses the effect of pileup on
all object-level observables at the HL-LHC. This suppression yields significant and
democratic improvements to many physics analyses -- including Higgs, BSM, LLPs, and di-Higgs physics --  by increasing signal
efficiencies or reducing the width of residual distributions for
discriminating variables.

The MIP timing detector is a conceptual realization  for such precision timing detector  at the HL-LHC. The MTD is composed of two subsystems based in different sensor technologies:  SiPMs+LYSO:Ce for the barrel region and LGADs for the endcap region. The technology choices were driven by performance, radiation, mechanics and schedule requirements and constraints. Both sensor technologies are mature and proven to achieve $\sim$\hspace{-0.01cm}30--50~ps time resolutions.

\bibliography{eprint}

\begin{thebibliography}{10}

\bibitem{Butler:2020886}
``{Technical Proposal for the Phase-II Upgrade of the CMS Detector},'' Tech.
  Rep. CERN-LHCC-2015-010. LHCC-P-008, CERN, Geneva, Jun 2015.

\bibitem{Butler:2055167}
``{CMS Phase II Upgrade Scope Document},'' Tech. Rep. CERN-LHCC-2015-019.
  LHCC-G-165, CERN, Geneva, Sep 2015.

\bibitem{Apollinari:2017cqg}
G.~Apollinari, O.~Brüning, T.~Nakamoto, and L.~Rossi, ``{High Luminosity Large
  Hadron Collider HL-LHC},'' {\em CERN Yellow Report}, p.~1, 2015.

\bibitem{CMS:2009nxa}
``Particle-flow reconstruction and global event description with the cms
  detector,'' {\em JINST}, vol.~12, p.~P10003, 2017.

\bibitem{Bertolini:2014bba}
D.~Bertolini, P.~Harris, M.~Low, and N.~Tran, ``{Pileup Per Particle
  Identification},'' {\em JHEP}, vol.~1410, p.~59, 2014.

\bibitem{DP2016_065}
{CMS Collaboration}, ``{Updates on Performance of Physics Objects with the
  Upgraded {CMS} detector for High Luminosity LHC},'' CMS Performance Note
  CMS-DP-2016-065, 2016.

\bibitem{gundacker2013time}
S.~Gundacker, E.~Auffray, B.~Frisch, P.~Jarron, A.~Knapitsch, T.~Meyer,
  M.~Pizzichemi, and P.~Lecoq, ``{Time of flight positron emission tomography
  towards 100~ps resolution with L(Y)SO: an experimental and theoretical
  analysis},'' {\em JINST}, vol.~8, p.~P07014, 2013.

\bibitem{LYSONIM}
D.~Anderson, A.~Apresyan, A.~Bornheim, J.~Duarte, C.~Pena, A.~Ronzhin,
  M.~Spiropulu, J.~Trevor, and S.~Xie, ``On timing properties of lyso-based
  calorimeters,'' {\em Nucl. Instrum. Meth. A}, vol.~794, p.~7, 2015.

\bibitem{Anderson:2015tia}
D.~Anderson {\em et~al.}, ``{Precision Timing Measurements for High Energy
  Photons},'' {\em Nucl.Instrum.Meth. A}, vol.~787, p.~94, 2015.

\bibitem{White:2014oga}
S.~N. White, ``{R\&D for a Dedicated Fast Timing Layer in the CMS Endcap
  Upgrade},'' {\em Acta Phys. Pol. B Proc. Suppl.}, vol.~7, p.~743, 2014.

\bibitem{Pellegrini201412}
G.~Pellegrini {\em et~al.}, ``{Technology developments and first measurements
  of Low Gain Avalanche Detectors (LGAD) for high energy physics
  applications},'' {\em Nucl. Instrum. Meth. A}, vol.~765, p.~12, 2014.

\bibitem{Cartiglia:2015iua}
N.~Cartiglia {\em et~al.}, ``{Design optimization of ultra-fast silicon
  detectors},'' {\em Nucl. Instrum. Meth. A}, vol.~796, p.~141, 2015.

\bibitem{Chatrchyan:2014fea}
S.~Chatrchyan {\em et~al.}, ``{Description and performance of track and
  primary-vertex reconstruction with the CMS tracker},'' {\em JINST}, vol.~9,
  p.~P10009, 2014.

\bibitem{deFlorian:2016spz}
D.~de~Florian {\em et~al.}, ``{Handbook of LHC Higgs Cross Sections: 4.
  Deciphering the Nature of the Higgs Sector},'' 2016.

\bibitem{CMS-PAS-EXO-14-007}
{CMS Collaboration}, ``{Enhanced scope of a Phase 2 CMS detector for the study
  of exotic physics signatures at the HL-LHC},'' CMS Physics Analysis Summary
  CMS-PAS-EXO-14-007, 2016.

\bibitem{Cerri:2018rkm}
O.~Cerri, S.~Xie, C.~Pena, and M.~Spiropulu, ``{Identification of Long-lived
  Charged Particles using Time-Of-Flight Systems at the Upgraded LHC
  detectors},'' 2018.

\bibitem{PhysRevD.94.113003}
A.~Coccaro, D.~Curtin, H.~J. Lubatti, H.~Russell, and J.~Shelton, ``Data-driven
  model-independent searches for long-lived particles at the {LHC},'' {\em
  Phys. Rev. D}, vol.~94, p.~113003, 2016.

\end{thebibliography}
\bibliographystyle{ieeetr}

\end{document}